\documentclass[12pt,a4paper,nohyper]{JHEP}
\usepackage{cite,epsfig}

\newcommand{\G}{\Gamma}

\newcommand{\cD}{{\cal D}}
\newcommand{\cO}{{\cal O}}
\newcommand{\wh}{\widehat}
\newcommand{\ep}{\epsilon}

\newcommand{\nn}{\nonumber}

\newcommand{\fm}{\mbox{\rm fm}}

\newcommand{\eqn}[1]{(\ref{#1})}

\newcommand{\mev}{\mbox{\rm MeV}}
\newcommand{\gev}{\mbox{\rm GeV}}

\newcommand{\MSb}{{\overline{MS}}}

\newcommand{\msmall}[1]{\mbox{$#1$}}
\newcommand{\tvs}{\vbox{\vskip 6mm}}
\newcommand{\cDt}{\widetilde{\cal D}}
\newcommand{\smvs}{\vbox{\vskip 8mm}}
\newcommand{\bmvs}{\vbox{\vskip 10mm}}

\newcommand{\gsim}{~{}_{\textstyle\sim}^{\textstyle >}~}

\newcommand{\vacl}{\langle 0|}
\newcommand{\vacr}{|0\rangle }
\newcommand{\prp}{\frac{1}{2}\,(1+\!\!\not{\!v})}

\newcommand{\quarkkond}{\langle \bar{q}q\rangle}
\newcommand{\gemkond}{\langle g\bar{q}\sigma G q\rangle}

\newcommand{\bhat}{\widehat{B}_T}

\title{The gauge invariant quark correlator\\
       in QCD sum rules and lattice QCD}

\author{H.G. Dosch, M. Eidem\"uller, M. Jamin\thanks{Heisenberg fellow}\\
Institut f\"ur Theoretische Physik, Universit\"at Heidelberg\\
Philosophenweg 16, 69120 Heidelberg, Germany\\
E--mail:
\email{H.G.Dosch@ThPhys.Uni-Heidelberg.DE}
\email{M.Eidemueller@ThPhys.Uni-Heidelberg.DE}
\email{M.Jamin@ThPhys.Uni-Heidelberg.DE} }

\author{E. Meggiolaro\\
Dipartimento di Fisica, Universit\`{a} di Pisa\\
Via Buonarroti 2, I--56127 Pisa, Italy\\
E--mail: \email{enrico.meggiolaro@sns.it} }

\abstract{Taking the gauge invariant quark correlator as an example, in this
work we perform a direct comparison of lattice QCD simulations with the QCD
sum rule approach. The quark correlator is first investigated in the
framework of QCD sum rules and the correlation length of the quark field
is calculated. Comparing the phenomenological part of the sum rule with
previous measurements of the quark correlator on the lattice, we are able
to obtain an independent result for the correlation length. From a fit of
the lattice data to the operator product expansion, the quark condensate
and the mixed quark--gluon condensate can be extracted.}

\preprint{HD--THEP--00--21\\
          IFUP--TH/2000--09}


\begin{document}


\section{Introduction}

Gauge invariant field correlators can serve as interesting examples for
studying non-perturbative aspects of QCD in the framework of lattice QCD
\cite{mm:94,rot:92} or QCD sum rules \cite{svz:79,nar:89,shi:92}. In addition,
they are a natural extension of the local condensates which appear in the sum
rule approach, due to the use of the operator product expansion (OPE)
\cite{w:69}. Phenomenological implications of nonlocal condensates have
previously been discussed in the literature \cite{gro:82,mr:86,mr:89,bbd:91,
r:91, br:91,mr:92}. The two most fundamental correlators are the gauge
invariant field strength correlator and the gauge invariant quark correlator.
The gluon field strength correlator has already been investigated in lattice
QCD \cite{gp:92,gmp:97,egm:97} and QCD sum rules \cite{ej:98,dej:99}. These
studies allowed to extract the correlation length of the gluon field.

On the other hand, the gauge invariant quark correlator has so far only been
measured on the lattice \cite{egm:98:1,egm:98:2}. In this work, we therefore
first present a QCD sum rule analysis of the gauge invariant quark correlator.
To this end, the quark correlator is related to a heavy--light meson
correlator \cite{shi:95,eid:97} in the heavy quark effective theory (HQET)
\cite{neu:94}. Thus the sum rule analysis will be related to previous
investigations of the same HQET correlators \cite{bg:92,bbbd:92} in the limit
of the heavy quark mass going to infinity. The correlation length of the quark
field is then given as the inverse of the binding energy of the light quark.

In the second, and more innovative, part of the present work, we perform a
direct comparison of the lattice data \cite{egm:98:1,egm:98:2} with the
representations of the correlation function in the QCD sum rule approach.
From the so--called phenomenological parametrisation --- a single resonance
plus the perturbative continuum as the simplest Ansatz --- we again extract
the correlation length of the quark field, in agreement with the direct
sum rule determination. Fitting the lattice data with the theoretical
correlator in the framework of the operator product expansion, we obtain
estimates of the quark and the mixed quark--gluon condensate which are in
reasonable agreement with sum rule phenomenology.

Our paper is organised as follows: In the next section we discuss the relation
of the gauge invariant quark correlator and the corresponding heavy quark
current correlator. In section 3 we set up the expressions needed for the
sum rule analysis and in section 4 we shall present our numerical results.
Section 5 compares our results with recent lattice determinations of the
quark correlator. In particular, a direct comparison of the OPE with the
lattice data will be performed. Finally, in section 6, we conclude with a
summary and an outlook.


\section{The gauge invariant quark correlator}

The central object of our investigation is the gauge invariant two--point
correlation function of quark fields,
\begin{equation}
\label{eq:2.1}
\cD_{\alpha\beta}(z) \equiv  \langle 0|T\{\bar{q}^a_\alpha(y)\,
\phi^{ab}(y,x)\,q^b_\beta(x)\}|0\rangle \,,
\end{equation}
where the string operator $\phi^{ab}(y,x)$ ensures the gauge invariance of
the correlator and is represented by
\begin{equation}
\label{eq:2.2}
\phi^{ab}(y,x) \equiv  [\,{\cal P}\,e^{igT^C
\int_0^1 d\lambda \,z^\mu\! A^C_\mu(x+\lambda z)}\,]^{ab} \,,
\end{equation}
with $z=y-x$. $T^C$ are the generators of SU(3) in the fundamental
representation and ${\cal P}$ denotes path ordering of the exponential. In
general, the gauge invariant field strength correlator could be defined using
an arbitrary gauge string connecting the end points $x$ and $y$, but in this
work we have restricted ourselves to a straight line, for it is only in this
case that the correlator \eqn{eq:2.1} is related to a heavy meson correlator
in HQET.

Using the path--integral formalism, we are able to derive a relation between
the correlator $\cD_{\alpha\beta}(z)$ and the correlator of a local, gauge
invariant current composed of a light quark field $q^a_\alpha(x)$ and an
infinitely heavy quark field $h^a_\alpha(x)$ \cite{eid:97}. Analogously to
HQET, the heavy quark field $h^a_\alpha(x)$ is constructed from the field
$Q^a_\alpha(x)$ with a finite mass $m_Q$ in the limit $m_Q\to\infty$,
\begin{equation}
\label{eq:2.3}
h^a_\alpha(x) \, = \, \lim_{m_Q\rightarrow\infty}\,\prp_{\alpha\beta}\,
e^{im_Qvx}\,Q^a_\beta(x) \,,
\end{equation}
where $v_\mu$ is the four--velocity of the heavy quark. In the case of
infinitely heavy quarks, the coupling to the gauge fields is given by the
effective HQET action $S_{eff}=\int d^4\!x \,\bar{h}\,iv^\mu D_\mu h$ with
$D_\mu=\partial_\mu-igA_\mu$ \cite{neu:94}. In the free field case the
heavy quark propagator is given by
\begin{eqnarray}
\label{eq:2.4}
S^{ab}_{\alpha\beta}(z) &\equiv& \vacl T\{h^a_\alpha(y)\,\bar{h}^b_\beta(x)\}
\vacr = \delta^{ab}\,\prp_{\alpha\beta}\,S(z) \nn \\
&=& \delta^{ab}\, \prp_{\alpha\beta}\, \frac{1}{v^0}\,\theta(z^0)\,
\delta\Big({\bf z}-\frac{z^0}{v^0}{\bf v}\Big) \,,
\end{eqnarray}
where  $v^0$ is the zero--component of the velocity. In addition, we have
the relation $v_\mu=z_\mu/|z|$ with $|z|\equiv\sqrt{z^2-i\ep}$ which follows
immediately from the constraints ${\bf z}=z^0{\bf v}/v^0$ and $v^2=1$.
In order to obtain a solution for the interacting case, we have to find
the inverse of the operator $iv^\mu D_\mu$. Using the following crucial
relation obeyed by the string operator,
\begin{equation}
\label{eq:2.5}
v^\mu \partial^y_\mu \,\phi(y,x) = v^\mu\,ig T^C A^C_\mu(y)\, \phi(y,x) \,,
\end{equation}
together with equation \eqn{eq:2.4} one can show that the solution for the
interacting case is found to be \cite{eid:97}
\begin{equation}
\label{eq:2.6}
\vacl T\{h^a_\alpha(y)\,\bar{h}^b_\beta(x)\,e^{iS_{eff}}\} \vacr  = 
S_{\alpha\beta}(z)\,\vacl [\,{\cal P}e^{igT^C \int_0^1 d\lambda\,
z^\mu\! A^C_\mu(x+\lambda z)}\,]^{ab} \vacr \,.
\end{equation}
The physical interpretation of this result is an infinitely heavy quark
moving from point $x$ to point $y$ with a four--velocity $v$, acquiring a
phase proportional to the path--ordered exponential. The limit $m_Q\to\infty$
is necessary in order to constrain the heavy quark to a straight line and
to decouple the spin interactions.

The result \eqn{eq:2.6} can be employed to relate the correlator \eqn{eq:2.1}
to correlators of gauge invariant local currents in HQET. To this end, we
define the pseudoscalar and scalar heavy meson currents as
\begin{equation}
\label{eq:2.7}
j_{P}(x) = \bar{q}^{a}_\alpha(x)\,(i\gamma_5)_{\alpha\beta}\,h_\beta^{a}(x) 
\quad \mbox{and} \quad
j_{S}(x) = \bar{q}^{a}_\alpha(x)\,h_\alpha^{a}(x) \,,
\end{equation}
as well as the correlators $\cDt_{\G}(z)$ with $\G=P$ or $S$:
\begin{equation}
\label{eq:2.8}
\cDt_{\G}(z) \equiv  \vacl T\{j_{\G}(y)\,j_{\G}^\dagger(x) \}\vacr \,.
\end{equation}
The two correlators could have also been defined with vector and axialvector
currents, but as it was shown in ref. \cite{bg:92}, the vector and axialvector
correlators are equal to the pseudoscalar and scalar correlators respectively
in the heavy mass limit. This result is related to the fact that in the heavy
mass limit the corresponding physical states become degenerate.
Integrating out the heavy quark fields in \eqn{eq:2.8}, we obtain the
expressions
\begin{eqnarray}
\label{eq:2.9}
\cDt_{P}(z) &=&  -\,\frac{1}{2}\,\vacl \bar{q}^a(y)(1-\!\!\not{\!v})
\phi^{ab}(y,x) q^b(x) \vacr \,S(z) \,, \nn \\
\cDt_{S}(z) &=&  \phantom{-}\,\frac{1}{2}\,\vacl \bar{q}^a(y)(1+\!\!\not{\!v})
\phi^{ab}(y,x) q^b(x) \vacr \,S(z) \,,
\end{eqnarray}
displaying the relation of the heavy meson correlators to the gauge invariant
quark correlator of equation \eqn{eq:2.1}.

In energy space the HQET correlator $\cDt_\G(w)$ obeys the spectral
representation 
\begin{equation}
\label{eq:2.12}
\cDt_\G(w) = \sum_k \frac{f_{\G,k}^2}{(E_{\G,k}-w-i\ep)} + \int
\limits_{w_0^\G}^\infty d\lambda\,\frac{\rho_\G(\lambda)}{(\lambda-w-i\ep)}\,,
\end{equation}
where $w=vq$ and the Fourier transform of the correlators $\cDt_\G(z)$ in
coordinate space is given by
\begin{equation}
\label{eq:2.13}
\cDt_{\G}(w) = i\! \int d^4\!z\, e^{iqz}\, \cDt_{\G}(z) \,.
\end{equation}
$E_{\G,k}$ represents the energy of the bound states and $f_{\G,k}$ is the
coupling of the state $k$ with quantum numbers $\G$ to the vacuum,
\begin{equation}
\label{eq:2.14}
\vacl j_{\G}(0)|H_{\G,k}\rangle = f_{\G,k} \,.
\end{equation}
The spectral densities are defined by $\rho_\G(\lambda)\equiv 1/\pi\,{\rm Im}\,
\cDt_\G(\lambda+i\ep)$, and finally $w_0^\G$ is the threshold energy of the
continuum contribution. Transforming the spectral representation \eqn{eq:2.12}
back to coordinate space, one obtains
\begin{eqnarray}
\label{eq:2.15}
\cDt_\G(z) & = & -i \int \frac{d^4\!q}{(2\pi)^4}\, e^{-iqz}\,\cDt_\G(w) 
\equiv S(z)\,\cD_\G(z) \nn \\
&=& S(z)\,\biggl\{\,\sum_k f_{\G,k}^2 e^{-iE_{\G,k}|z|} +
\int\limits_{w_0^\G}^\infty d\lambda\, \rho_\G(\lambda)\, e^{-i\lambda |z|}\,
\biggr\} \,.
\end{eqnarray}
The factorisation of the heavy quark propagator and the relations \eqn{eq:2.9}
ensure a representation of the gauge invariant quark correlators $\cD_\G(z)$
as given by the expression inside the curly brackets. Let us already remark
that the correlator decays as a simple exponential in the Euclidean region.
The correlation length will therefore be given by the inverse of the lowest
lying bound state energy.


\section{The sum rules}

We now turn to the theoretical side of the sum rules which arises from 
calculating the correlator of equation \eqn{eq:2.8} in the framework of the 
operator product expansion \cite{svz:79,w:69}. The perturbative contributions
for the spectral density up to next--to--leading order in the strong coupling
constant have been calculated in \cite{bg:92}. Up to $\cO(m^2)$ in the light
quark mass, they are found to be:
\begin{eqnarray}
\label{eq:3.1}
\rho_\G^{pt}(w) &=& w^2 \sum\limits_{n=0}^{2}\Big[\,d_{n0}^\G+a(d_{n1}^\G+
d_{n1L}^\G L)\,\Big]\biggl(\frac{m}{w}\biggr)^n \nn \\
\smvs
&=& \frac{N_c}{8\pi^2}\,\Biggl\{\,
w^2 \biggl[\,4+C_F a\Big(17+\msmall{\frac{4}{3}}\pi^2-6L\Big)\,\biggr] \\
\smvs
&& \pm\,wm\biggl[\,4+C_F a\Big(24+\msmall{\frac{4}{3}}\pi^2-12L\Big)\,\biggr]
-\,m^2 \biggl[\,2+C_F a\Big(3-9L\Big)\,\biggr]\,\Biggr\} \nn
\end{eqnarray}
where $a\equiv\alpha_s/\pi$ and $L\equiv\ln(2w/\mu)$. Equation \eqn{eq:3.1}
also defines the coefficients $d_{ni}^\G$ which will be utilised below. As
above and in all the following, the upper sign corresponds to the
pseudoscalar and the lower sign to the scalar current.

In order to suppress contributions in the dispersion integral coming from
higher exited states and from higher dimensional operators, it is convenient
to apply a Borel transformation $\wh{B}_T$ with $T$ being the Borel variable.
The Borel transformation also removes the subtraction constants which are
present in the dispersion relation satisfied by the correlators. Some useful
formulae for the Borel transformation are collected in the appendix. For the
phenomenological side of the sum rules, equation \eqn{eq:2.12}, we only take
the lowest lying resonance and approximate the spectral density by the
perturbative expression \eqn{eq:3.1}, assuming quark--hadron duality for
$w>w_0^\G$. We then obtain
\begin{equation}
\label{eq:3.2}
\wh{\cD}_\G(T) \, = \, f_\G^2 \,e^{-E_\G/T} +
\int\limits_{w_0^\G}^\infty d\lambda\,\rho^{pt}_\G(\lambda)\,e^{-\lambda/T} \,.
\end{equation}
Let us remark that equation \eqn{eq:3.2} takes exactly the same form as the
expression of \eqn{eq:2.15} inside the curly brackets, which represents the
gauge invariant quark correlator, if we identify $1/T$ with the Euclidean
space-time coordinate.

The perturbative contribution relevant for the sum rules is the full
correlator minus the corresponding continuum contribution:
\begin{equation}
\label{eq:3.3}
\wh{\cD}_\G^{pt}(T) - \wh{\cD}^{co}_\G(T,w_0^\G) \, = \,
\int\limits_0^\infty d\lambda\,\rho^{pt}_\G(\lambda)\,e^{-\lambda/T} -
\int\limits_{w_0^\G}^\infty d\lambda\,\rho^{pt}_\G(\lambda)\,e^{-\lambda/T} \,,
\end{equation}
where $\wh{\cD}^{co}_\G(T,w_0^\G)$ denotes the perturbative continuum part.
After the Borel transformation the correlators satisfy homogeneous
renormalisation group equations. Thus we can improve the perturbative
expressions by resumming the logarithmic contributions. With the help of
the following general integral formula \cite{jm:95},
\begin{equation}
\label{eq:3.4} \int
\limits_{w_0^\G}^\infty d\lambda\,\lambda^{\alpha-1}\ln^n \frac{2\lambda}{\mu}
\,e^{-\lambda/T} \,=\, T^\alpha \sum_{k=0}^{n} {n\choose k} \ln^k\frac{2T}{\mu}
\left[\frac{\partial^{n-k}}{\partial\alpha^{n-k}}\,\Gamma\!\left(\alpha,
\frac{w_0^\G}{T}\right)\right] \,,
\end{equation}
the perturbative contribution to the sum rules is found to be: 
\begin{eqnarray}
\label{eq:3.5}
\lefteqn{ \wh{\cD}^{pt}_\G(T)-\wh{\cD}^{co}_\G(T,w_0^\G) \, = \, } \\
\smvs
&& T^3\Biggl(\!\frac{a(2T)}{a(\mu)}\!\Biggr)^{\gamma_1/\beta_1}
\sum\limits_{n=0}^2 \phi(3\!-\!n,y)\Biggl[\,d_{n0}^\G+a\Biggl(d_{n1}^\G+
d_{n1L}^\G\frac{\phi'(3\!-\!n,y)}{\phi(3\!-\!n,y)}\Biggr)\Biggr]\Biggl(\!
\frac{m(2T)}{T}\!\Biggr)^n \nn
\end{eqnarray}
where $\phi(\alpha,y) \equiv \Gamma(\alpha)-\Gamma(\alpha,y)$,
$\phi'(\alpha,y) \equiv \partial/\partial\alpha\,\phi(\alpha,y)$ and
$y=w_0^\G/T$. Some explicit expressions for the incomplete $\Gamma$-function
$\Gamma(\alpha,y)$ and the functions $\phi(\alpha,y)$, $\phi'(\alpha,y)$
can also be found in the appendix. In our notation $\beta_1=(11N_c-2N_f)/6$
is the first coefficient of the QCD $\beta$-function. The anomalous dimension
$\gamma_1$ of both currents is easily found by reexpanding the running coupling
and mass in terms of $a(\mu)$ and $m(\mu)$. The resulting expression is
\begin{equation}
\label{eq:3.6}
\gamma_1 \, = \, -\,\frac{d_{n1L}^\G}{d_{n0}^\G} - n\,\gamma_{1m}
\, = \, \frac{3}{2}\,C_F \, = \, 2 \,,
\end{equation}
where the first coefficient of the mass anomalous dimension,
$\gamma_{1m} = 3C_F/2$, has been used. The $\mu$-dependence of the correlators
is due to the fact that the pseudoscalar and scalar currents are not
renormalisation group invariant quantities. However, the product of $m$ times
the current, $m\!\cdot\! j_\G$, is renormalisation group invariant in the
full theory. Taking into account the additional multiplicative renormalisation
of the heavy quark current in HQET \cite{bg:92,eh:90}, one again finds
$\gamma_1= \gamma_{1m}$.

Essential for QCD sum rule analyses are the non--perturbative contributions
coming from vacuum condensates. In the operator product expansion, the
correlation function is expanded in powers of $1/w$ corresponding to higher
and higher dimensional condensates. In our case the dimension three condensate
$\langle\bar{h}h\rangle$ vanishes since the heavy quark mass is infinite.
After Borel transformation and up to operators of dimension five, the
non--perturbative contribution takes the form \cite{bg:92}:
\begin{equation}
\label{eq:3.7}
\wh{\cD}^{np}_\G(T) \, = \, \mp\,\frac{\quarkkond_\mu}{2}\,\biggl[\,1 \mp
\frac{m}{4T} + \frac{3}{2}\,C_F a\,\biggr] \pm \frac{\gemkond_\mu}{32 T^2} \,.
\end{equation}
The next condensate contribution would be of dimension six. We have checked
explicitly in our numerical analysis that this contribution to the sum rule
is small. Thus we shall neglect all condensate contributions for dimensions
higher than five in this work. For consistency, we have also omitted the known
order $\cO(a^2)$ contribution to the quark condensate \cite{bg:92}, because
the corresponding corrections to the perturbative part have not yet been
calculated. We have, however, verified that also this correction only has
a minor impact on our numerical results.

In the case of the non--perturbative part, the scale dependence of the
correlator is implicit in the $\mu$-dependence of the quark condensate.
Indeed, again $m\quarkkond$ is scale independent and therefore $\quarkkond$
scales inversely like the quark mass yielding the same $\mu$-dependence as for
the perturbative contribution. For the mixed quark--gluon condensate, the
next--to--leading order corrections have not been calculated and thus in the
numerical analysis below, we shall neglect its scale dependence.


\section{Numerical analysis}

After equating the phenomenological and the theoretical contributions to the
correlation functions, we end up with the following sum rule:
\begin{equation}
\label{eq:4.1}
K_\G(T) \, \equiv \, f_\G^2\, e^{-E_\G/T} \, = \,
\wh\cD^{pt}_\G(T)-\wh{\cD}^{co}_\G(T,w_0^\G) + \wh\cD^{np}_\G(T) \,.
\end{equation}
The binding energy $E_\G$ can be obtained by dividing the derivative of the
sum rule \eqn{eq:4.1} with respect to $-1/T$ by the original sum rule:
\begin{eqnarray}
\label{eq:4.2}
E_\G \, &=& \, -\,\frac{\partial}{\partial(1/T)}\,\ln K_\G  \nn \\ 
\smvs
&=& -\,\frac{\frac{\partial}{\partial(1/T)}\Big(\wh\cD^{pt}_\G(T) -
\wh\cD^{co}_\G(T,w_0^\G) + \wh{\cD}^{np}_\G(T)\Big)}{\Big(\wh\cD^{pt}_\G(T) -
\wh\cD^{co}_\G(T,w_0^\G) + \wh{\cD}^{np}_\G(T) \Big)} \,.
\end{eqnarray}
Analogously to equation \eqn{eq:3.5} an expression for the derivative of
the perturbative contribution can be obtained with the help of formula
\eqn{eq:3.4}:
\begin{eqnarray}
\label{eq:4.3}
\lefteqn{-\frac{\partial}{\partial(1/T)}\Big(\wh{\cD}^{pt}_\G(T)-
\wh{\cD}^{co}_\G(T,w_0^\G)\Big) \, = \, } \\
\smvs
&& T^4\Biggl(\!\frac{a(2T)}{a(\mu)}\!\Biggr)^{\gamma_1/\beta_1}
\sum\limits_{n=0}^2 \phi(4\!-\!n,y)\Biggl[\,d_{n0}^\G+a\Biggl(d_{n1}^\G+
d_{n1L}^\G\frac{\phi'(4\!-\!n,y)}{\phi(4\!-\!n,y)}\Biggr)\Biggr]\Biggl(\!
\frac{m(2T)}{T}\!\Biggr)^n \nn \,.
\end{eqnarray}
The corresponding derivative of the non--perturbative contribution is easily
calculated from equation \eqn{eq:3.7}. Because the renormalisation of the
correlators is multiplicative, it is clear that the binding energy $E_\G$
is a physical quantity in the sense that it is scale and scheme independent.
On the contrary, this is not the case for the decay constant $f_\G$, as we
shall discuss in more detail below.

\FIGURE{\epsfig{file=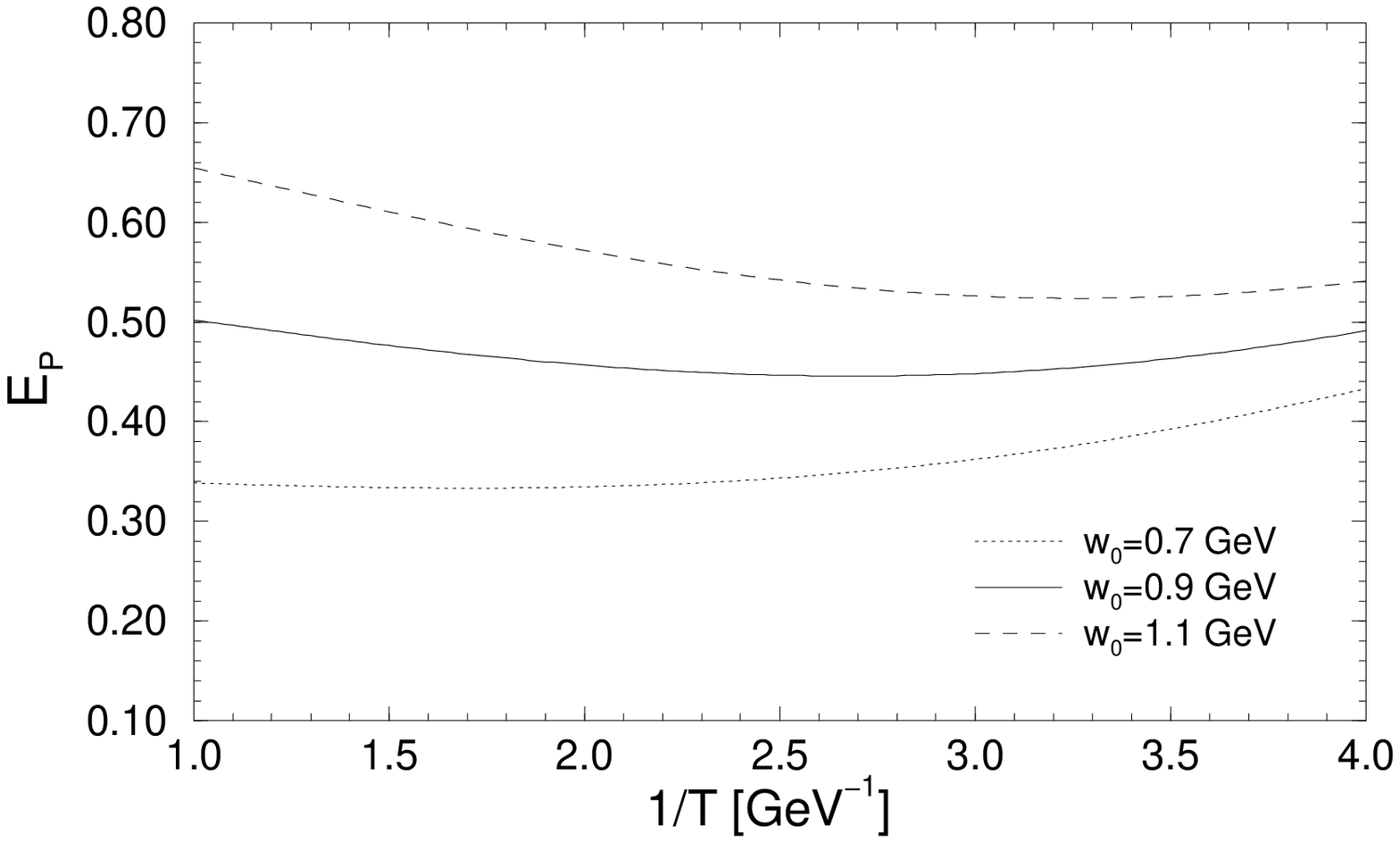,width=12cm}
\caption{Pseudoscalar binding energy as a function of $1/T$ for different
values of the continuum threshold $w_0^P$.}
\label{fig1}}
Let us first consider the pseudoscalar correlator for a vanishing light
quark mass $m_q=0$. As the central values for our input parameters we use 
$\quarkkond (1\,\gev)=-(235\pm 20\,\mev)^3$ for the quark condensate,
$\gemkond=M_0^2\,\quarkkond$ with $M_0^2=(0.8\pm 0.2)\,\gev^2$ \cite{bi:82,
nar:88,op:88} for the mixed condensate and $\Lambda^{(3)}_{\MSb}=340\,\mev$.
The latter value corresponds to three light quark flavours and $\alpha_s(M_Z)=
0.119$. In principle, the coupling constant in the next--to--leading order term
could be evaluated at any scale $\mu$. As our central value in the numerical
analysis we have chosen $\mu=1.2\,\gev$ but we shall investigate the
variation of $\mu$ below. In figure~\ref{fig1}, we display the pseudoscalar
binding energy $E_P$ as a function of $1/T$ for different values of the
continuum threshold $w_0^P$. A sum rule window, were both the continuum
as well as the condensate contributions are reasonably small, can be found
around $1/T=2.5-3.0\,\gev^{-1}$. As can be seen from figure~\ref{fig1}, best
stability is achieved for $w_0^P\approx 0.9\,\gev$. Estimating the value of
$E_P$ in the given range $0.7\,\gev<w_0^P<1.1\,\gev$, we obtain:
\begin{equation}
\label{eq:4.4}
E_P \,=\, 450 \pm 100 \,\mev  \qquad (m_q \,=\, 0) \,.
\end{equation}

\FIGURE{\epsfig{file=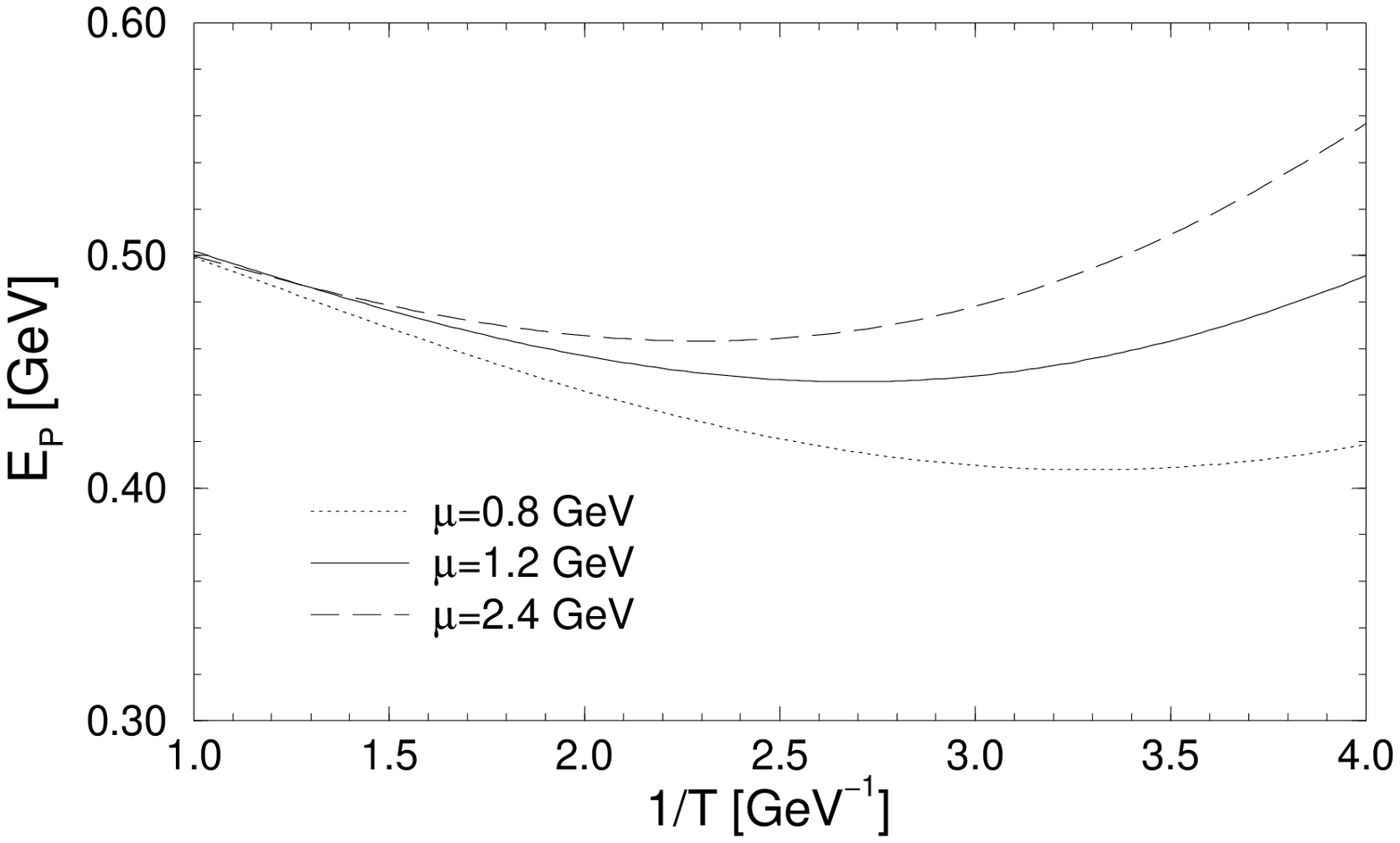,width=12cm}
\caption{Pseudoscalar binding energy as a function of $1/T$ for
$w_0^P=0.9\,\gev$ and different values of the renormalisation scale $\mu$.}
\label{fig2}}
Although the next--to--leading order QCD corrections to the sum rule of
equation~\eqn{eq:4.1} are very large, of the order of 100\%, in the ratio
of equation~\eqn{eq:4.3} they cancel to a large extent. To investigate the
sensitivity of our result to higher order corrections, we next study the
dependence on the renormalisation scale $\mu$.
The dependence of $E_P$ on the renormalisation scale $\mu$ is shown in
figure~\ref{fig2} for $w_0^P=0.9\,\gev$ and $\mu=0.8\,\gev$, $1.2\,\gev$
and $2.4\,\gev$. One should not take $\mu$ smaller than $0.8\,\gev$
because then also the radiative corrections to the quark condensate
become unacceptably large. We observe that inspite of the huge corrections
to the correlation function, the scale dependence of $E_P$ is relatively mild.
Adding this uncertainty to the error on $E_P$, our central result for $E_P$
is found to be
\begin{equation}
\label{eq:4.5}
E_P \,=\, 450 \pm 150 \,\mev  \qquad (m_q \,=\, 0) \,.
\end{equation}
The additional uncertainty coming from the errors on the input parameters
is small compared to the estimated uncertainty and can be neglected.

\FIGURE{\epsfig{file=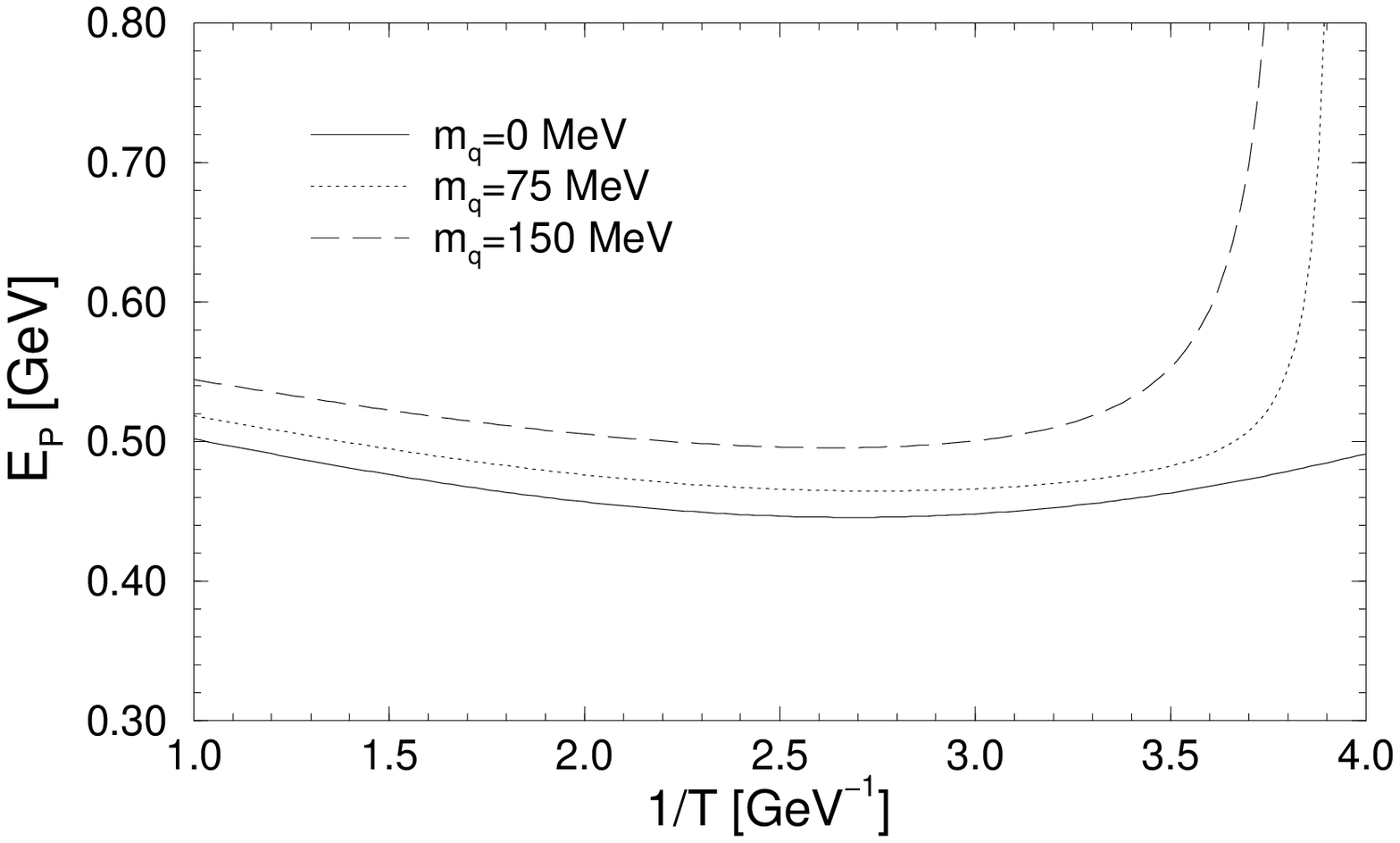,width=12cm}
\caption{Pseudoscalar binding energy as a function of $1/T$ for
$w_0^P=0.9\,\gev$, $\mu=1.2\,\gev$ and different values of the light quark
mass $m_q$. The absolute value of $\quarkkond$ has been reduced in accord
with the quark mass.}
\label{fig3}}
Let us now investigate the influence of a finite light quark mass $m_q$.
In figure~\ref{fig3}, we have plotted $E_P$ as a function of $1/T$ for
$w_0^P=0.9\,\gev$, $\mu=1.2\,\gev$ and three values of the light quark
mass $m_q$, namely $m_q=0\,\mev$, $75\,\mev$ and $150\,\mev$. The last
value is in the region of the mass of the strange quark and the intermediate
value is interesting for comparison with lattice QCD results. It is well known
from QCD sum rules that at the mass of the strange quark, the absolute
value of the corresponding quark condensate is reduced by roughly 30\%
\cite{nar:89}. This reduction has been applied for obtaining the dashed curve
in figure~\ref{fig3}. Lacking further information, for the dotted curve
corresponding to $m_q=75\,\mev$ the quark condensate has been reduced by
10\%.

Qualitatively, we find that the binding energy increases with increasing
light quark mass and decreasing quark condensate. For $m_q=150\,\mev$ this
increase turns out to be of the order of $50\,\mev$. Such a value is only
about half of the mass splitting of $B$ and $B_s$ mesons in the $B$-meson
system as well as $D$ and $D_s$ mesons in the $D$-meson system which is of
the order of $100\,\mev$. Further comparison with the heavy meson systems
will be presented below. The instability for $1/T\gsim 3.5\,\gev^{-1}$ results
from the running quark mass $m_q(2T)$ which explodes in this region due to
uncontrollably large higher order corrections.

\FIGURE{\epsfig{file=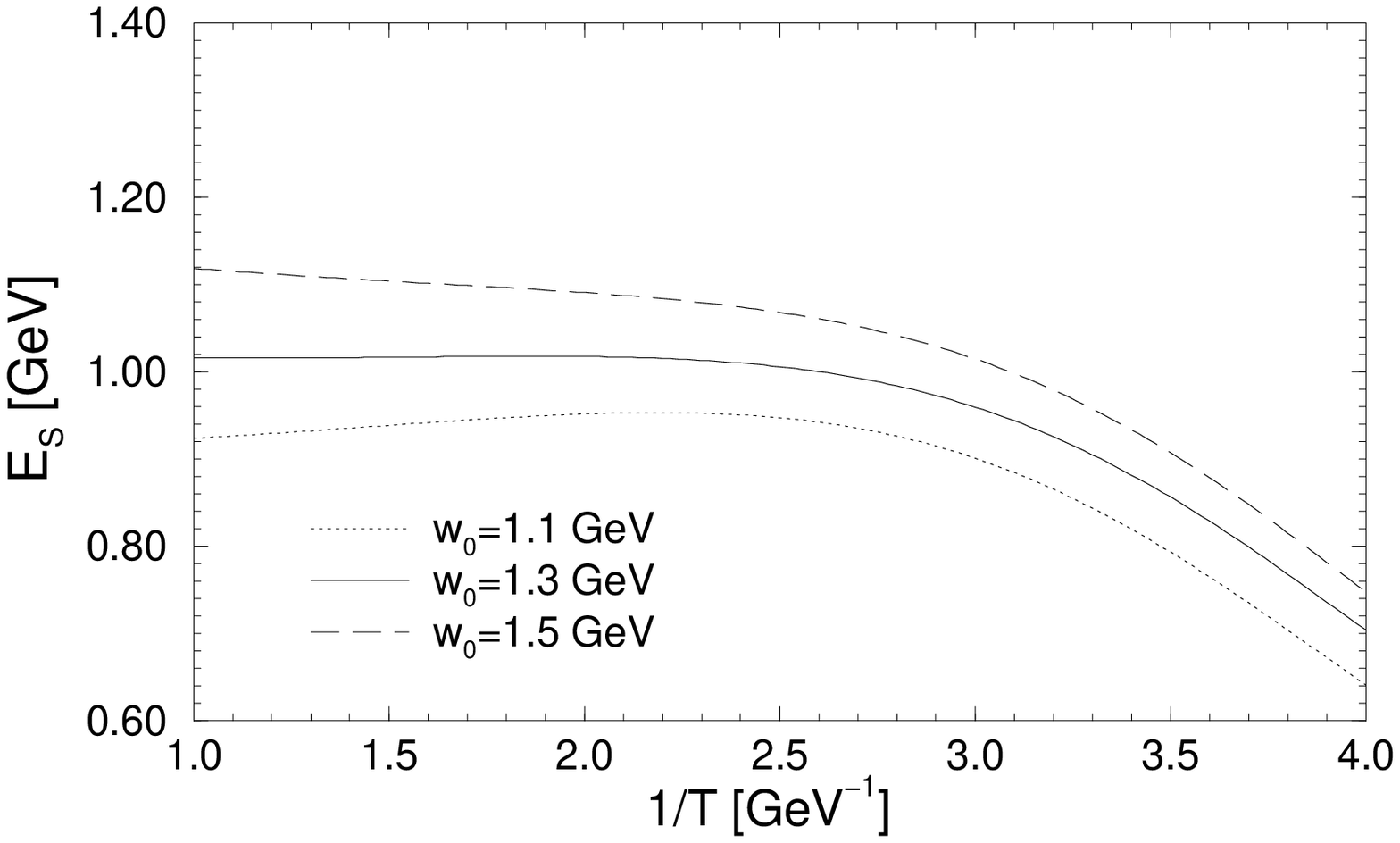,width=12cm}
\caption{Scalar binding energy as a function of $1/T$ for different
values of the continuum threshold $w_0^S$.}
\label{fig4}}
Let us next turn to the scalar correlator. In figure~\ref{fig4} the scalar
binding energy is shown for three different values of the continuum
threshold $w_0^S$. We observe that generally the stability of the scalar sum
rule as not as good as in the pseudoscalar case. Here, the region of best
stability is around $1/T=2.0\,\gev^{-1}$. Adding the uncertainty from the
scale dependence, the scalar binding energy is found to be:
\begin{equation}
\label{eq:4.8}
E_S \,=\, 1.0 \pm 0.2 \,\gev  \qquad (m_q \,=\, 0) \,.
\end{equation}
Compared to the pseudoscalar correlator, the dependence of $E_S$ on the
light quark mass $m_q$ is somewhat stronger. Approximately, we obtain
$E_S(m_q)\approx E_S(0)+m_q$ for $m_q\leq 150\,\mev$. However, in the scalar
case for increasing quark mass and decreasing quark condensate, the sum
rule becomes less stable and thus we refrain from making more quantitative
statements.

To conclude this section, let us comment on the decay constants $f_\G$.
As has been already discussed above, because of the renormalisation of the
heavy meson current, $f_\G$ depends on the renormalisation scale and scheme.
Therefore, it is convenient to define renormalisation-group invariant decay
constants $\wh f_\G$. At the next--to--leading order, $\wh f_\G$ takes the
form \cite{bg:92,bbbd:92}:
\begin{eqnarray}
\label{eq:4.9}
\wh f_\G \, = \, f_\G\, \alpha_s(\mu)^{\gamma_1/2\beta_1}
(\,1+K\,a\,) \quad \mbox{with} \quad
K \, = \, \frac{5}{12}-\frac{285-7\pi^2}{54\beta_1}+\frac{107}{8\beta_1^2} \,.
\end{eqnarray}
As central values, from the sum rule of equation \eqn{eq:4.1} we then find
$\wh f_P=0.3\,\gev^{3/2}$ and $\wh f_S=0.5\,\gev^{3/2}$. Since the
next--to--leading QCD correction to the correlators is very large, this is
also true for the uncertainty on $\wh f_\G$. Thus we shall not dicuss the
heavy meson decay constants further.


\section{Comparison with lattice results}

In ref. \cite{egm:98:1} the two following quark--antiquark nonlocal
condensates have been determined on the lattice:
\begin{eqnarray}
\label{eq:5.1}
C_0(x) &=& -\displaystyle\sum_{f=1}^4\, \langle {\rm Tr} [ \bar{q}^f_\alpha(0)
\phi(0,x) q^f_\alpha(x) ] \rangle \,, \nn \\
C_v(x) &=& -\,{x_\mu \over |x|} \displaystyle\sum_{f=1}^4\, \langle {\rm Tr}
[ \bar{q}^f_\alpha(0)(\gamma^\mu_E)_{\alpha\beta} \phi(0,x) q^f_\beta(x) ]
\rangle \,.
\end{eqnarray}
The trace in \eqn{eq:5.1} is taken with respect to the colour indices and
$f$ is a flavour index for the light quarks. $\gamma_E^\mu$ are Euclidean
Dirac matrices, defined as:
$\gamma_{E4} = \gamma^0$, $\gamma_{Ei} = -i \gamma^i$ $(i = 1,2,3)$. Other
insertions of gamma matrices than those containing $1$ and $\gamma^\mu_E$
vanish by P and T invariance.

In order to avoid confusion with the correlators introduced in section~2
the correlators $C_0(x)$ and $C_v(x)$ are called respectively ``spin--zero''
nonlocal condensate and ``longitudinal--vector'' nonlocal condensate.
They are simply related to the correlators $\cDt_\G$ of equation \eqn{eq:2.9}
if the latter are continued to Euclidean space time:
\begin{eqnarray}
\label{eq:5.2}
\cDt_{P}(x) &=& \frac{1}{2N_f}\,\Big(C_v(x)+C_0(x)\Big)\,S(x) \,, \nn \\
\cDt_{S}(x) &=& \frac{1}{2N_f}\,\Big(C_v(x)-C_0(x)\Big)\,S(x) \,,
\end{eqnarray}
where $N_f$ is the number of quark flavours.

The lattice computations of ref. \cite{egm:98:1} have been performed both in
the {\it quenched} approximation and in full QCD using the SU(3) Wilson
action for the pure--gauge sector and four degenerate flavours of
{\it staggered} fermions, so that the sum over the flavour index $f$ goes
from 1 to 4. In full QCD the nonlocal condensates have been measured on a
$16^3 \times 24$ lattice at $\beta = 5.35$ and two different values of the
quark mass: $a \cdot m_q = 0.01$ and $a \cdot m_q = 0.02$ ($a$ being the
lattice spacing). For the {\it quenched} case the measurements have been
performed on a $16^4$ lattice at $\beta = 6.00$, using valence quark masses
$a \cdot m_q = 0.01$, $0.05$, $0.10$, and at $\beta = 5.91$ with a quark
mass $a \cdot m_q = 0.02$. Further details, as well as a remark about the
reliability of the results obtained for the longitudinal--vector nonlocal
condensate, can be found in \cite{egm:98:1}. The scalar functions $C_0$ and
$C_v$ introduced in ref. \cite{egm:98:1} have a more complicated spectral
representation than $\cDt_{P}$ and $\cDt_{S}$, since they receive contributions
both from scalar and pseudoscalar intermediate states. Nevertheless, the
correlator $C_0$ has the advantage of not receiving perturbative contributions
in the chiral limit $m_q\rightarrow 0$ with $m_q$ being
the mass of the light quark:
\begin{equation}
\label{eq:5.3}
\lim_{x\rightarrow 0} \ C_0(x) \rightarrow \frac{N_f N_c m_q}{\pi^2 x^2}
\qquad \mbox{and} \qquad
\lim_{x\rightarrow 0} \ C_v(x) \rightarrow \frac{2 N_f N_c}{\pi^2 x^3} \,,
\end{equation}
where $N_c$ is the number of colours.

Since the available lattice results for $C_v (x)$ are not conclusive, in
what follows we shall concentrate on the spin--zero nonlocal condensate
$C_0(x)$. In ref. \cite{egm:98:1} a best fit to the data with the following
function has been performed:
\begin{equation}
\label{eq:5.4}
C_0(x) \, = \, A_0 \exp (-\mu_0 x) + {B_0 \over x^2} \,.
\end{equation}
The perturbative--like term $B_0/x^2$ takes the form obtained in the leading
order in perturbation theory, if the chiral limit $m_q\to 0$ (see equation
\eqn{eq:5.3}) is approached. Using the ansatz \eqn{eq:5.4}, the correlation
length $\lambda_0 \equiv 1/\mu_0$ of the spin--zero quark--antiquark nonlocal
condensate can be extracted. At the lightest quark mass $a \cdot m_q = 0.01$
in full QCD one obtains the value $\lambda_0 = 0.63^{+0.21}_{-0.13}$ fm
\cite{egm:98:1} corresponding to $\mu_0 = 310\pm 80\,\mev$. Within errors
this value agrees with the value for $E_P$ obtained from the sum rule analysis.

Some observations are, however, necessary at this point. One should point
out that the parametrisation \eqn{eq:5.4} is a sort of ``hybrid''
parametrisation, since it contains a ``perturbative'' term $B_0/x^2$, which
should reproduce the behaviour predicted by perturbation theory at short
distances, and a ``non--perturbative'' term $A_0 \exp (-\mu_0 x)$, which
should reproduce the predicted exponential behaviour at large distances. 
A simple exponential term is not dominant in the range of physical distances
where lattice data are taken, i.e., $x \approx 0.2 \div 0.8$ fm. This could
shed some doubt on the identification of $\mu_0 = 1/\lambda_0$ with the
binding energy $E_P$, as determined in section~4.

Nevertheless, at the distances where lattice data are taken the operator
product expansion should still be a reasonable approximation. One can therefore
try to fit directly the OPE expression to the lattice data. In this case
however a running coupling would lead to a Landau pole around 1 fm and one
should confine the comparison to distances small compared to this scale.
On the other hand most treatments of non--perturbative QCD are based on the
assumption that at large distances only non--perturbative effects prevail.
This can be achieved by freezing the coupling at a certain value for distances
larger than a critical one. Given the fact that we have only few data points
we consider here only the leading order terms in the OPE. In addition, the
correlator is scheme dependent and to perform a consistent comparison between
the lattice data and the OPE at the next--to--leading order, a perturbative 
calculation in the lattice scheme would be necessary.

At the leading order the spin--zero correlator $C_0(x)$ up to operators of
dimension seven is given by:
\begin{equation}
\label{eq:5.5}
C_0(x)/N_f \, = \, \frac{N_c m_q^2}{\pi^2 x}\,K_1(m_q x) - \Big[\,1+ \msmall{
\frac{1}{8}}\,m_q^2 x^2\,\Big]\quarkkond + \frac{x^2}{16}\,\gemkond \,,
\end{equation}
where in the perturbative part through the Bessel function $K_1(z)$ we have
kept all orders in the quark mass. Fits for the quark and the mixed condensate
from the different sets of lattice data of ref. \cite{egm:98:1} are given in
table~\ref{tab1}. The presented errors just correspond to the statistical
$1\,\sigma$ errors resulting from the fit if the $\chi^2/d.o.f.$ is normalised
to one. $m_L$ is the lattice mass converted to physical units and $m_q$ the
mass appearing in the OPE of equation \eqn{eq:5.5}. Since we work at the
leading order, the scale and scheme dependence of the quark mass are not
under control.
\TABLE{
\begin{tabular}{|c|c|c|c|c|c|}
\hline
$\beta$ & $a\cdot m$ & $m_L$ [MeV] & $m_q$ [MeV] & $-\quarkkond^{1/3}$ [MeV] &
$M_0^2$ [GeV$^2$] \\
\hline
6.00, q & 0.01 &  19 & $ 38\pm 4$ & $244\pm 4$ & $0.47\pm0.03$ \\
5.91, q & 0.02 &  33 & $ 73\pm11$ & $198\pm12$ & $0.44\pm0.06$ \\
6.00, q & 0.05 &  96 & $150\pm14$ & $186\pm15$ & $0.68\pm0.09$ \\
6.00, q & 0.10 & 192 & $276\pm22$ & $175\pm20$ & $0.87\pm0.12$ \\
5.35, f & 0.01 &  20 & $ 28\pm 2$ & $168\pm 4$ & $0.47\pm0.04$ \\
5.35, f & 0.02 &  33 & $ 54\pm 6$ & $177\pm 9$ & $0.45\pm0.04$ \\
\hline
\end{tabular}
\caption{Determination of the condensates by direct comparison of lattice
data with the OPE. q denotes quenched approximation, f full QCD with 4 light
fermions, $m_L$ is the input lattice mass and $m_q$ the OPE mass appearing
in equation \eqn{eq:5.5}.\label{tab1}} }

We note that the condensates come out with the correct order of magnitude.
Qualitatively also the decrease of the quark condensate with increasing quark
mass is found albeit somewhat stronger than expected from phenomenology
\cite{nar:89}. A direct extraction of the quark condensate from the uncooled
values of the spin--zero quark correlator \cite{egm:98:1} at zero distance
led to the value $\quarkkond(1\,\gev) = -\,(235\pm 15\,\mev)^3$ in perfect
agreement with phenomenological determinations. We have again parametrised
the mixed quark--gluon condensate through $\gemkond=M_0^2 \quarkkond$. The
dependence of $M_0^2$ on the quark mass is controversial in the literature
\cite{djn:89,bd:92}.  From our fits we obtain an increase of $M_0^2$ with
increasing quark mass, supporting the findings of ref. \cite{djn:89}.

The non--perturbative part of the OPE \eqn{eq:5.5} can be expressed at short
distances by the Gaussian $-\quarkkond\exp(-M_0^2 x^2/16)$ which at large
distances displays an exponential falloff. Fitting the lattice data for the
quenched calculation with $a\cdot m_q = 0.01$ to the perturbative part of
\eqn{eq:5.5} and the Gaussian non--perturbative contribution, we find
$m_q=33\pm 2\,\mev$, $-\quarkkond^{1/3}=254\pm 2\,\mev$ and
$M_0^2=0.74\pm0.02\,\gev^2$. This value for $M_0^2$ is more compatible
with phenomenological determinations \cite{bi:82,nar:88,op:88}. However, the
result shows that higher order corrections in the OPE have some importance
and it gives an indication about the systematic uncertainties.

Another way to improve the hybrid expansion \eqn{eq:5.4} consists in taking
the higher resonances into account in the same way as it is done in the
sum rule technique as developed by SVZ \cite{svz:79}, namely by including 
the perturbative continuum above a threshold $w_0$. In leading order for the
spin--zero nonlocal condensate this yields:
\begin{eqnarray}
\label{eq:5.6}
C_0(x)/N_f &=& f_{P}^2\,e^{-E_{P} x} -f_{S}^2\,e^{-E_{S} x}\nn \\
&&+\frac{N_c}{2 \pi^2 x^3}\,\Big((2 + 2 w_0^{P}x + (w_0^{P}x)^2)e^{-w_0^{P}x}-
(2 + 2 w_0^{S}x + (w_0^{S}x)^2)e^{-w_0^{S}x}\Big)\nn \\
&&+\frac{N_c m_q}{2 \pi^2 x^2}\,\Big((1+w_0^{P}x)e^{-w_0^{P}x}+
(1+w_0^{S}x)e^{-w_0^{S}x}\Big) \,.
\end{eqnarray}
\FIGURE{\epsfig{file=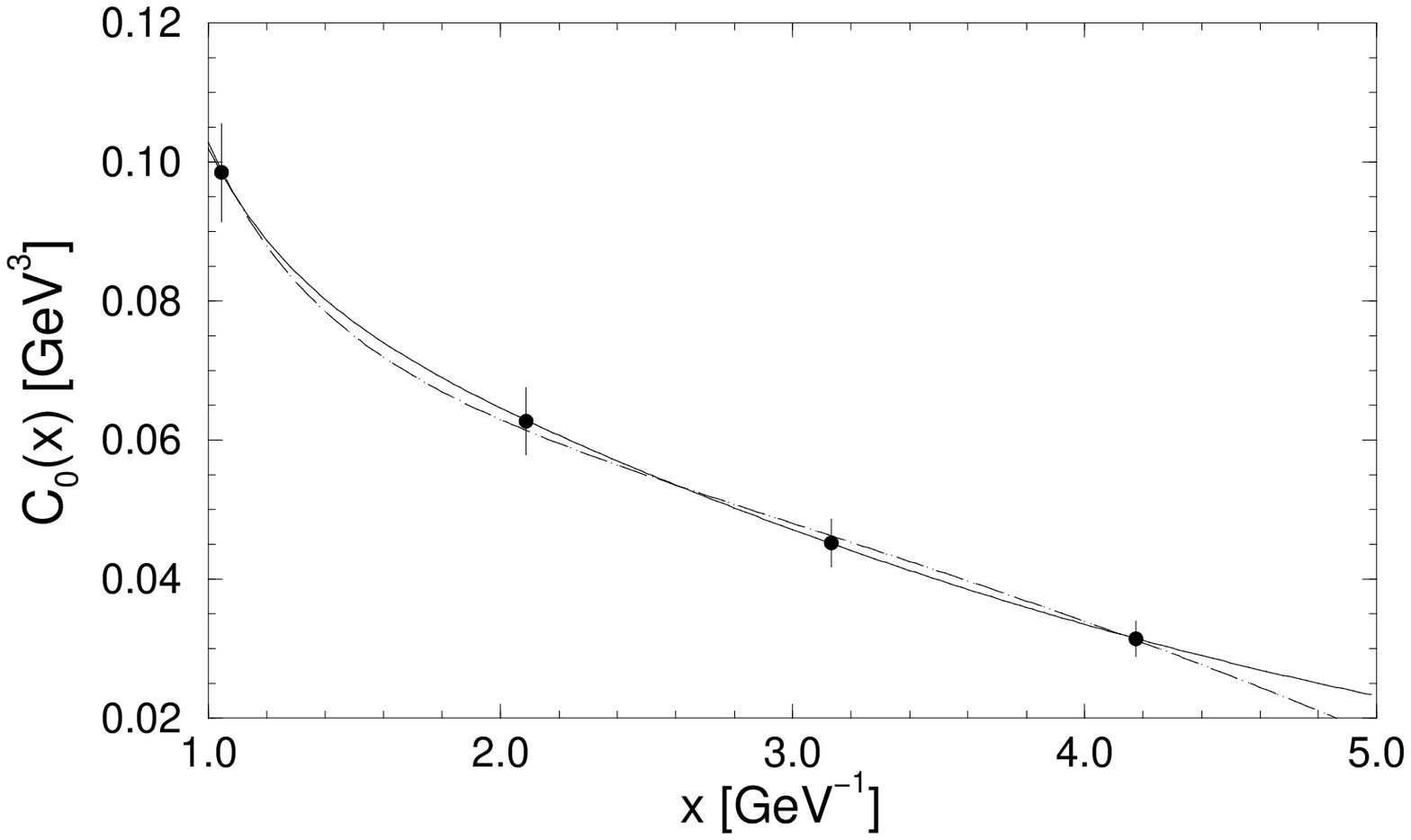,width=12cm}
\caption{Lattice data from \cite{egm:98:1} with $a\cdot m_q = 0.01$, quenched
calculation. Dashed curve: OPE with condensates of table 1, first row, solid
curve: equation \eqn{eq:5.6} with $f_{P}=0.245\,\gev^{3/2}$, $E_{P}=411\,\mev$,
$w_0^P=1.0\,\gev$, $f_S=0$ and $w_0^S=0$.\label{fig5}} }
In figure 5 we have displayed the lattice data for the quenched calculation
with $a\cdot m_q = 0.01$. The dashed curve is the fit with the OPE \eqn{eq:5.5}
and the parameters given in the  first row of table 1. The solid curve is a
fit with equation \eqn{eq:5.6} where for simplicity, and lacking enough data
points, we have neglected the scalar resonance by setting $f_S=0$ and
$w_0^S=0$. This means that in the scalar channel only the perturbative term
was taken into account and no resonance was singled out. Using in addition
$m_q=38\,\mev$, the fit leads to $f_{P}=0.245\pm 0.002\,\gev^{3/2}$,
\begin{equation}
\label{eq:5.7}
E_P \,=\, 411 \pm 3 \,\mev  \,,
\end{equation}
and $w_0^{P}=1.0\pm 0.01\,\gev$. Within the uncertainties, these results
are in good agreement to the sum rule determination of the last section.

In other lattice investigations \cite{All:91,Ale:91} of the heavy meson
systems the main interest was to determine the leptonic decay constant of
the heavy light meson in HQET. In both cases it was found that in order to
isolate the ground state contribution one has in some way to suppress the
higher state contributions. This was done by two different smearing
mechanisms. Though the authors claimed that on the lattice the mass gap
$E_P$ is not a physical quantity and therefore divergent in the continuum
limit they determined this quantity for their lattice spacing. The difference
of the mass gaps $E_S-E_P$ however is a physical quantity and must therefore
have a continuum limit. In table~\ref{tab2} we collect the results for these
quantities for the different approaches and compare to our findings presented
above.
\TABLE{
\begin{tabular}{|l|c|c|c|c|}
\hline
Ref.& $\beta $ & $a^{-1}$ [GeV] & $E_P$ [GeV] & $E_S - E_P$ [GeV] \\
\hline
\cite{Ale:91}   & 5.74 & 1.53 & 1.93* & 0.69 \\
\cite{Ale:91}   & 6.0  & 2.03 & 1.58* & 0.49 \\
\cite{All:91}   & 6.0  & 2    & 1.3*  & -    \\
\cite{egm:98:1} & 6.0  & 1.9  & 0.31  & -    \\
this work L     & 6.0  & 1.9  & 0.41  & -    \\
this work SR    & -    &  -   & 0.45  & 0.55 \\
\hline
\end{tabular}
\caption{Pseudoscalar mass $E_P$ and difference between scalar and
pseudoscalar mass in different approaches for different values of the strong
coupling $\beta=2N_c/g_s^2$ and lattice spacing $a$. This work L is an
analysis of the data lattice data from \cite{egm:98:1} according to equation
\eqn{eq:5.6}, this work SR the sum rule analysis of section 4. In the lattice
calculations of \cite{All:91,Ale:91} the pseudoscalar mass is not defined
for $a\to 0$.\label{tab2}} }


\section{Conclusions}

In this paper we have investigated the gauge invariant quark correlator.
We have set up a relation between the correlator and a corresponding 
correlator of gauge invariant currents in the heavy quark effective theory.
The relevant currents are interpolating fields of pseudoscalar and scalar
heavy--light mesons. With the method of QCD sum rules and cooled lattice data
it is possible to extract the mass gap between the heavy quark pole mass
and the lightest bound states of each current. This mass gap is a physical
quantity in the sense that it is scale and scheme independent in perturbation
theory to all orders in the strong coupling constant. The resulting values
have been found to be $E_P= 450\pm 150\,\mev$ for the pseudoscalar state and
$E_S=1.0\pm 0.2\,\gev$ for the scalar state. The corresponding correlation
lengths are: $a_P=0.44^{+0.22}_{-0.11}\,\fm$ and $a_s=0.20^{+0.05}_{-0.03}\,
\fm$. For the cooled lattice data of \cite{egm:98:1,egm:98:2} an analysis of
a perturbative term plus an exponential gave the correlation length
$a_P=0.63^{+0.21}_{-0.13}\,\fm$; the modified analysis with a single resonance
plus continuum of equation \eqn{eq:5.6} yields the value $E_P= 411\,\mev$,
corresponding to $a_P=0.48\,\fm$.

These results can be compared to the spectrum of heavy pseudoscalar mesons.
Let us make the very simple assumption that up to corrections of $1/m_Q$ where
$m_Q$ is the heavy quark pole mass, the heavy meson mass is equal to the
heavy quark mass plus binding energy:
\begin{equation}
\label{eq:4.6}
E_P - \frac{c}{m_Q} \, = \, M_P - m_Q \,.
\end{equation}
Here $M_P$ is the mass of the pseudoscalar meson and $c$ is a constant.
Assuming in addition that this relation is valid for both the $B$ and
$D$ mesons, we can solve for $E_P$ and $c$. Using $m_b=5.0\pm0.2\,\gev$,
$m_c=1.8\pm0.2\,\gev$ and experimental values for $M_B$ and $M_D$,
as central values for $E_P$ and $c$ we obtain:
\begin{equation}
\label{eq:4.7}
E_P \, = \, 400 \, \mev \qquad \mbox{and} \qquad
c \, = \, 0.6 \, \gev^2 \,.
\end{equation}
The result for $E_P$ is in good agreement with the other determinations
presented above. Nevertheless, we should remark that the latter values are
very sensitive to the heavy quark pole mass and with the estimated errors
on $m_b$ and $m_c$, the uncertainty on $E_P$ is of the order of 100\%. In
addition, from the value of $c$ we see that the assumption of small $1/m_Q$
corrections is not valid in the charm case. Still, we find the very qualitative
agreement of our results with the spectrum of pseudoscalar mesons noteworthy.

For lattice simulations of the gauge invariant quark correlator where the
higher states were suppressed by a smearing procedure the mass gaps are
said to diverge in the continuum limit. Though there is no indication of
such a divergence with increasing $\beta$ (see table~2) the very large values
for $E_P$ found in \cite{All:91} and \cite{Ale:91} are definitely not physical.
The difference $E_S-E_P$ is however a physical quantity. It has been estimated
in \cite{Ale:91} and the values listed in table~\ref{tab2} agree with the
result from the sum rules better than expected in view of the large errors.

The lattice data for $C_0(x)$ have directly been compared to the operator
product expansion in leading order QCD. The results for the continuum values
of the quark mass, the quark and the mixed condensates are well compatible
with the values determined from other sources. The results from the quenched
approximation are nearer to the continuum values than those from full QCD. The
known decrease of the modulus of quark condensate with the quark mass is also
confirmed by the lattice calculations. The ratio of the mixed to the quark
condensate $M_0^2$ comes out to be the same in the quenched and full QCD.
There is some controversy on the dependence of $M_0^2$ on the quark mass
\cite{djn:89,bd:92}. The lattice data support an increase of $M_0^2$ with
the quark mass which was found in \cite{djn:89}. For distances smaller
than approximately 1 fm where the operator product expansion can still be
applied a good fit to the non--perturbative part of the correlator is also
given by the Gaussian $-\quarkkond\exp(-M_0^2 x^2/16)$ which has previously
been used in the nonlocal sum rule approach \cite{r:91,br:91,mr:92}.
 
In our opinion a direct comparison of lattice QCD simulations with QCD sum
rules in the framework of the operator product expansion opens a novel route
to augment our knowledge on low--lying hadronic states and non--perturbative
QCD. Taking the results of this work as encouraging, we intend to further
pursue this approach in the future.


\acknowledgments

M.~Eidem\"uller thanks the Landesgraduiertenf\"orderung at the University of
Heidelberg for financial support. M.~Jamin would like to thank the Deutsche
Forschungsgemeinschaft for support.

 
\newpage

\appendix

\section{Appendix}

The Borel transformation in HQET is defined as
\begin{equation}
\label{eq:a.1}
\bhat \equiv \lim_{-w,n\rightarrow \infty} \frac{(-w)^{n+1}}{\Gamma(n+1)}
\left(\frac{d}{dw}\right)^n ,\qquad T=\frac{-w}{n}>0\quad \mbox{fixed} \,.
\end{equation}
Using this definition, a central formula for the Borel transformation is
found to be
\begin{equation}
\label{eq:a.2}
\bhat \ \frac{1}{(E-w-i\ep)^{\alpha}} = \frac{1}{\Gamma(\alpha) T^{\alpha-1}}
\,e^{-E/T} \,.
\end{equation}

Below, we have collected some analytic formulae for the incomplete Gamma
function and the functions $\phi(\alpha,y)$ and $\phi'(\alpha,y)$ defined as
\begin{eqnarray}
\label{eq:a.3}
\phi(\alpha,y) &\equiv& \Gamma(\alpha)-\Gamma(\alpha,y) \,, \nn \\
\phi'(\alpha,y) &\equiv& \frac{\partial}{\partial \alpha}
\Big(\Gamma(\alpha)-\Gamma(\alpha,y)\Big) \,,
\end{eqnarray}
which are helpful for the numerical analysis of the sum rules.
\begin{eqnarray}
\label{eq:a.4}
\Gamma(2,y)&=& e^{-y} (1+y) \nn\\
\tvs
\Gamma(3,y)&=& e^{-y} (2+2y+y^2) \nn\\
\smvs
\Gamma'(2,y)&=& \Gamma(0,y)+e^{-y} [1+(1+y)\ln y]  \nn\\
\tvs
\Gamma'(3,y)&=& 2\Gamma(0,y)+e^{-y} [3+y+(2+2y+y^2)\ln y] \nn\\
\bmvs
\phi(n,y) &=& \Gamma(n) \left(1-e^{-y}\sum_{k=0}^{n-1} \frac{y^k}{k!}\right)
\,, \qquad n=1,2,...\nn\\
\bmvs
\phi'(\alpha,y) &=& \int_0^y dt\ \ln t \ e^{-t} \, t^{\alpha-1} \nn\\
\smvs
\phi'(1,y) &=& -\gamma_E-\Gamma(0,y)-e^{-y}\,\ln y \nn\\
\tvs
\phi'(2,y) &=& 1-\gamma_E-\Gamma(0,y)-e^{-y}\left(1+(1+y)\ln y\right) \nn\\
\tvs
\phi'(3,y) &=& 3-2\gamma_E-2\Gamma(0,y)
-e^{-y}\left(3+y+\left(2+2y+y^2\right)\ln y\right) \nn\\
\tvs
\phi'(4,y) &=& 11-6\gamma_E-6\Gamma(0,y)\nn\\
& &-e^{-y}\left(11+5y+y^2+\left(6+6y+3y^2+y^3\right)\ln y\right) \,.
\end{eqnarray}


\newpage
\providecommand{\href}[2]{#2}\begingroup\raggedright\endgroup

\end{document}